# Composed multicore fiber structure for direction-sensitive curvature monitoring

Joel Villatoro 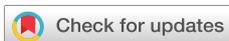 ; Josu Amorebieta; Angel Ortega-Gomez; Enrique Antonio-Lopez; Joseba Zubia 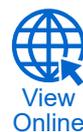 ;
Axel Schülzgen 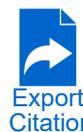 ; Rodrigo Amezcua-Correa

CrossMark

View Online

Export Citation

# Composed multicore fiber structure for direction-sensitive curvature monitoring

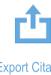

View Online    Export Citation    CrossMark

Joel Villatoro,[1,2,a] Josu Amorebieta,[1] Angel Ortega-Gomez,[1] Enrique Antonio-Lopez,[3] Joseba Zubia,[1] Axel Schülzgen,[3] and Rodrigo Amezcua-Correa[3]

## AFFILIATIONS

[1] Department of Communications Engineering, University of the Basque Country UPV/EHU, Torres Quevedo Plaza 1, E-48013 Bilbao, Spain
[2] IKERBASQUE—Basque Foundation for Science, E-48011 Bilbao, Spain
[3] CREOL, The College of Optics & Photonics, University of Central Florida, P.O. Box 162700, Orlando, Florida 32816-2700, USA

[a] Author to whom correspondence should be addressed: agustinjoel.villatoro@ehu.eus

## ABSTRACT

The present work deals with a curvature sensor that consists of two segments of asymmetric multicore fiber (MCF) fusion spliced with standard single mode fiber (SMF). The MCF comprises three strongly coupled cores; one of such cores is at the geometrical center of the MCF. The two segments of MCF are short, have different lengths (less than 2 cm each), and are rotated 180° with respect to each other. The fabrication of the sensor was carried out with a fusion splicing machine that has the means for rotating optical fibers. It is demonstrated that the sensor behaves as two SMF–MCF–SMF structures in series, and consequently, it has enhanced sensitivity. The device proposed here can be used to sense the direction and amplitude of curvature by monitoring either wavelength shifts or intensity changes. In the latter case, high curvature sensitivity was observed. The device can also be used for the development of other highly sensitive sensors to monitor, for example, vibrations, force, pressure, or any other parameter that induces periodic or local curvature or bending to the MCF segments.

## INTRODUCTION

Multicore fibers (MCFs) are revolutionary waveguides[1,2] that have multiple individual cores sharing a common cladding. In general, MCFs have diameters similar to that of a standard telecommunications optical fiber. The cores of an MCF can be well isolated from each other to avoid interactions between them. In this manner, each core behaves as an independent waveguide. Completely the opposite is also possible; this means that the cores can be in close proximity to each other to allow coupling between them. In the latter case, the fiber is called coupled-core MCF and supports supermodes.[3]

The unique features of MCFs provide new alternatives for the development of innovative devices whose functionalities cannot be easily achieved with conventional optical fibers. For example, ultra-thin lensless endoscopes[4] for biomedical applications and minimal intrusive shape sensors have been demonstrated.[5,6] MCFs with coupled cores offer also new possibilities for the development of simple and compact devices that can be used to monitor vibrations and bending,[7,8] among other parameters.

With regard to fiber optic curvature sensors, so far, a variety of configurations based on conventional fibers have been proposed and demonstrated (see Refs. 9–14). However, to the best of the authors' knowledge, such curvature sensors have not reached high readiness level. This suggests that it is important to investigate new alternatives to devise functional fiber optic curvature sensors.

MCFs with isolated cores offer multiple alternatives to build curvature sensors. For example, curvature sensors based on interferometers,[15–18] twisted MCFs,[19] or directional couplers[20] have been demonstrated. Some drawbacks of these sensors are the need of bulk optics to interrogate them, their insensitivity to the direction of curvature, their fragility as, in some cases, the MCF must be tapered, and the high insertion losses. Strongly coupled MCFs with quasi-symmetric core distribution have also been demonstrated for direction-insensitive curvature sensing.[21,22]

MCFs with a series of Bragg gratings[23–25] or long period gratings[26–28] in some or in all the cores can also be used to sense curvature. In fact, MCF curvature sensors based on Bragg gratings have reached a commercial level, but their high cost may limit their use to high-end applications. Some disadvantages of grating-based MCF curvature sensors include complex fabrication and expensive interrogation. Moreover, the curvature on some MCFs with gratings can induce coupling between cores. Such coupling can induce errors in the measurements of curvature.

Fiber optic curvature sensors have potential applications in shape sensing,[6,14] that is why they have attracted considerable research interest in recent years. Ideally, a fiber optic curvature sensor must be cost effective and must provide the amplitude and the direction of curvature. In addition, the sensor must be sensitive, simple, reliable, and very small in diameter, so it can be integrated to devices, instruments, or structures. We believe that the fiber optic curvature sensors reported to date cannot provide all these desirable characteristics.

Here, we propose a highly sensitive curvature sensor based on a strongly coupled MCF. Our device is easy to fabricate and requires a simple (low cost) interrogation system. In addition, our sensor is able to provide the amplitude and direction of curvature even by monitoring intensity changes. To achieve the curvature sensor with the aforementioned features, we used two short segments of different lengths of an MCF that comprises three identical cores. The two MCF segments are fusion spliced and rotated 180° with respect to each other and are inserted in a conventional single mode fiber.

The structure reported here can also be used to devise other sensors to monitor any parameter that induces point or periodic curvature to the MCF. Some examples may include force, pressure, and vibration sensors or accelerometers.

## SENSOR FABRICATION AND WORKING MECHANISM

In Fig. 1(a), we show the cross section of the MCF used to fabricate the sensor. The fiber has three coupled cores made of germanium-doped silica embedded in a cladding made of pure silica. The diameter of each core is approximately 9 $\mu$m, and the cores are separated from each other by 11 $\mu$m approximately. It can be noted that one core is at the geometrical center of the MCF. The numerical aperture of each core of the MCF is identical to that of an SMF (0.14). Due to the matching between the numerical apertures of both

fibers, the insertion losses of our devices are low as demonstrated previously.[7,21]

The architecture of our curvature sensor is shown in Fig. 1(b). Such a structure is fabricated by fusion splicing two segments of different lengths (typically less than 20 mm each) of the aforementioned MCF with a conventional SMF. The two segments of MCF are rotated 180° with respect each other; the reason of this angle is explained below. A reflector or mirror at the distal end of the SMF allows the sensor to operate in reflection mode, which has the advantages described in the following.

The fabrication of the device shown in Fig. 1(b) can be carried out with a splicing machine that has means of rotating optical fibers. In our case, we used a specialty fiber splicer (a Fujikura FSM-100P+) in which an *ad hoc* splicing program was implemented. With such a program, the end face of the two segments of MCF was inspected to orient the cores before the splicing. In all cases, the splices were carried out with a cladding alignment method. Under such splicing conditions, the cores located in the geometrical center of the two segments of MCF and the unique core of the SMF were axially aligned and permanently joined together. The two segments of MCF were intentionally rotated 180° to achieve an SMF–MCF1–MCF2–SMF structure in which the two cores outside the center of the MCFs were upward in one part of the structure and downward in the other part. We will see that such a structure behaves as a dual supermode coupler in series.

To understand the working mechanism of the device shown in Fig. 1(b), we carried out simulations based on the finite difference method with commercial software (FimmWave and FimmProp by Photon Design) and different experiments. In Fig. 2, we show the propagation of two different wavelengths from the lead-in SMF to the lead-out SMF in an SMF–MCF1–MCF2–SMF structure with the dimensions described in the figure. It can be seen that at 1500 nm, the guided light does not reach the lead-out SMF. On the other hand, light at 1550 nm propagates with losses. Consequently, in the referred structure, maximum transmission can be expected at 1550 nm and minimum at 1500 nm.

In addition to the simulations, we analyzed mathematically our device by considering that it is composed of two parts. Let us consider first the case when $L_2 = 0$. In this case, we will have an SMF–MCF1–SMF structure. To predict the transmission intensity of such a structure, we have to consider the following situations: (i) The three cores of the MCF are identical, i.e., they have the same diameter and the same refractive index; (ii) the distance between the MCF cores is the same; (iii) the central core of the MCF is excited with the fundamental SMF mode, and (iv) the MCF is composed by evanescently coupled single-mode cores. In our case, the latter assumptions are valid in the 1200 nm–1600 nm wavelength range. Under these conditions, two supermodes are excited in the MCF. Such supermodes have non-zero intensity in the central core of the MCF.[8]

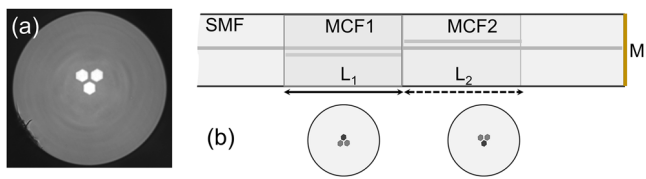

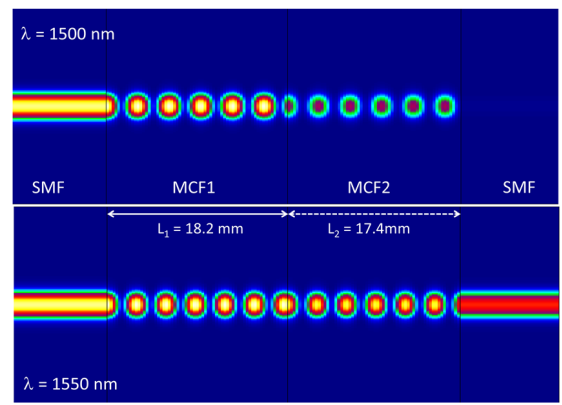

**FIG. 1.** (a) Micrograph of the MCF used to fabricate the samples. (b) Drawing of the device in which the two segments of MCF are rotated 180° with respect to each other. $L_1$ and $L_2$ are the lengths of the segments MCF1 and MCF2, respectively, and M is the mirror.

**FIG. 2.** Simulations of light propagation in an SMF–MCF1–MCF2–SMF structure. The following values were considered: $L_1$ = 12.20 mm and $L_2$ = 11.40 mm. The analyzed wavelengths are indicated.

The transfer function of the SMF–MCF1–SMF structure can be calculated by means of the coupled mode theory.[29] The transfer function is a periodic function of wavelength ($\lambda$) and can be expressed as[30–32]

$$I_{1T}(\lambda, L_1) = 1 - (2/3)\sin^2\left(\sqrt{3}\pi\Delta n L_1/\lambda\right). \quad (1)$$

In Eq. (1), $\Delta n$ is the effective refractive index difference between the two excited supermodes. $\Delta n$ depends on the wavelength, refractive index, dimensions, and separation between the cores of the MCF. For the MCF shown in Fig. 1(a), $\Delta n$ was found to be $4.66 \times 10^{-4}$. Now, if $L_1 = 0$, we will have an SMF–MCF2–SMF structure of length $L_2$. The transfer function of such a structure can also be expressed by Eq. (1), but with $L_2$ instead of $L_1$.

Let us now calculate the transfer function of an SMF–MCF–SMF structure when the SMF at the final extreme has a reflector or mirror on its face [see Fig. 1(b)]. In this case, the structure can be considered as two SMF–MCF–SMF structures in series. As demonstrated by several groups, the transfer function of two parallel fiber devices placed in series is the product of the individual transfer functions.[33–36] Thus, if a single SMF–MCF–SMF structure with $L_1$ (or $L_2$) is interrogated in reflection, the transfer function is simply $I_{1R} = I_{1T}^2$ (or $I_{2R} = I_{2T}^2$).

If the device shown in Fig. 1(b) is excited with a broadband source, the reflection measured with a photodetector or spectrometer will be

$$R(\lambda) = I_s(\lambda)\left[I_{1T}(\lambda, L_1)I_{2T}(\lambda, L_2)\right]^2. \quad (2)$$

In Eq. (2), $I_s(\lambda)$ is the spectral power distribution of the excitation light source. In a practical situation, such a light source can be a narrow-band light emitting diodes (LED) whose spectral distribution is Gaussian.

## RESULTS AND DISCUSSION

The interrogation of the device depicted in Fig. 1(b) is simple. In our case, we used a superluminescent light emitting diode (SLED) with peak emission at 1550 nm and a FWHM of 60 nm as the light source, a conventional fiber optic coupler (or circulator), and a photodetector or a miniature spectrometer (Ibsen I-MON-512) connected by a universal serial bus (USB) cable to a personal computer. Unless otherwise stated, in all our experiments, the cleaved end of the SMF segment after the MCF2 was used as a reflector. The reflectivity in this case was less than 4%.

In Fig. 3, we show the normalized reflection spectra of SMF–MCF–SMF structures in three different cases. The plots with dotted and dashed lines correspond to the spectra of individual structures with $L_1 = 12.20$ mm and $L_2 = 11.40$ mm. As the lengths of the MCF segments are short, the periods of the reflection spectra are long, and thus, it is not possible to observe two consecutive maxima in the monitored wavelength range. The shadowed area beneath the solid line represents the reflection spectrum observed when two segments of MCF, one with $L_1 = 12.20$ mm and the other with $L_2 = 11.40$ mm, were spliced together, but one segment of MCF was rotated 180° with respect to the other. The reflection spectrum of the SMF–MCF1–MCF2–SMF structure coincides with the spectrum that is obtained when the spectra shown in dotted and dashed lines are multiplied and then normalized. It can be noted that the experimental results shown in Fig. 3 agree with the simulations described in Fig. 2. Therefore, we can conclude that the reflection of the device depicted in Fig. 1(b) can be calculated with Eq. (2) as it can be treated as two SMF–MCF–SMF structures in series.

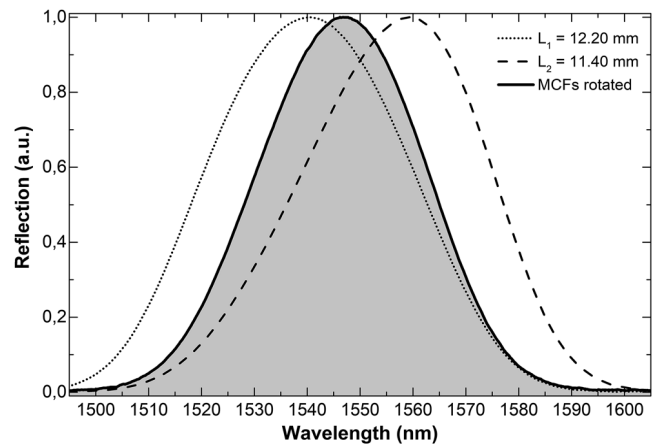

**FIG. 3.** Reflection spectra observed when the structure is SMF–MCF–SMF in which the lengths of MCFs are 12.20 mm (dashed line) and 11.4 mm (dotted line). The shadowed area beneath the solid line is the reflection spectrum observed when a 12.20 mm-long and an 11.40 mm-long segment of MCF are fusion spliced and rotated 180° with respect to each other.

To assess the performance of our composed MCF device as a curvature sensor, we carried out simulations, which are summarized in Fig. 4. In the figure, we show the reflection spectra of an SMF–MCF1–MCF2–SMF structure built with $L_1 = 17.4$ mm and $L_2 = 18.2$ mm at different values of curvature. It was assumed that the structure was bent in the MCF1–MCF2 junction and that both segments of MCF experienced the same curvature. The curvature was assumed to be applied in four different directions with respect to the orientations of the MCF cores. Any other orientation of the cores with respect to curvature will be contained between the four cases shown in Fig. 4. From the simulations, it can be concluded that the reflection intensity of our device will increase or decrease depending on the direction of curvature. This means that our device can distinguish the amplitude and direction of curvature.

To corroborate the above predictions, a simple setup, schematically shown in Fig. 5, was implemented. The SMF segments were secured with two fiber chucks that were mounted on respective rotators (HFR001 from Thorlabs). The chuck rotators were separated by a fixed distance and were secured on an optical breadboard that was placed in a vertical position. A fiber chuck was used as a mass (20 g) to keep the tension of the fibers constant. The measurements of curvature were carried out at different orientations of the MCFs, between 0 and 180° in steps of 30°, with respect to curvature [see Fig. 5]. A translation stage with micrometer resolution was used to bend the structure in a controlled manner. The stage bent the device close to the MCF1–MCF2 junction. The value of curvature ($C$) on the device was calculated with the following equation: $C = 12h/d^2$ (see Ref. 12), where $h$ is the displacement of the translation stage and $d$ is the separation between the two fiber rotators.

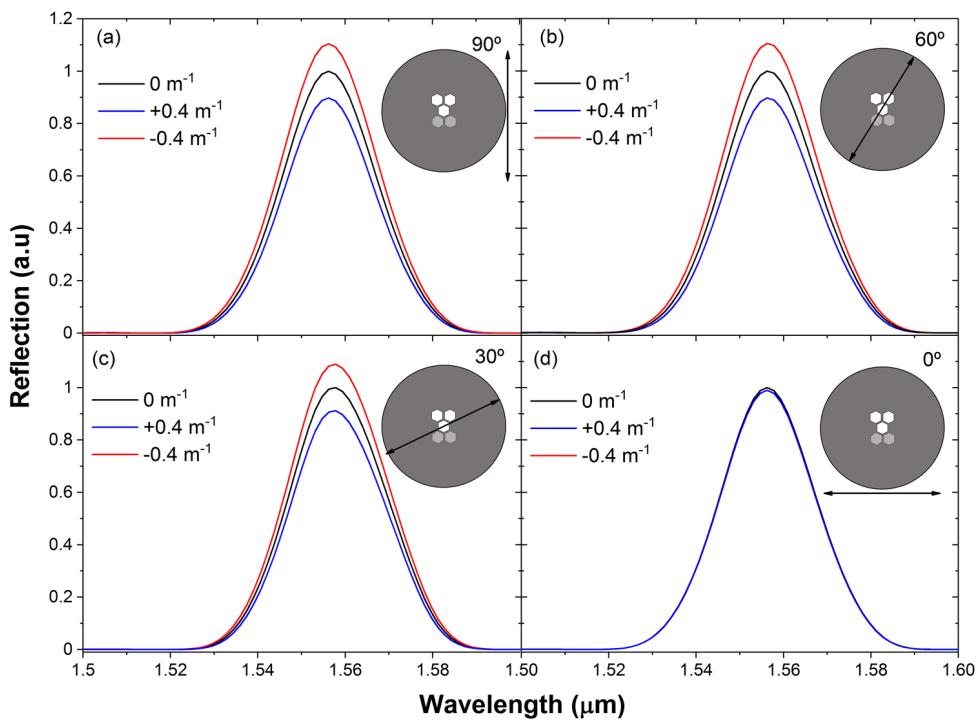

In the setup described in the above paragraph, any displacement of the translation stage (or change of $h$) causes bending to the two segments of MCF. However, the effect on them was different as the cores outside the center of the MCF had a different position with respect to the applied curvature. As demonstrated in Ref. 8, the asymmetric MCF used here is highly sensitive to bending. In addition, the direction of the bending can be distinguished when the MCF cores are oriented properly. Therefore, high sensitivity to curvature and capability to distinguish the direction of curvature were expected with an SMF–MCF1–MCF2–SMF structure. For this reason, we fabricated the structure as shown in Fig. 1(b) with the cores of the MCF1 and MCF2 segments rotated 180° with respect to each other.

A device fabricated with a segment of 17.4 mm of MCF fusion spliced to another segment of 18.2 mm was characterized in detail. As mentioned before, the cores of the MCF segments were in opposite orientation. Wavelength shifts and intensity changes were monitored at each value of curvature. In the former case, a spectrometer was used, while in the latter case, a low cost InGaAs photodiode (S154C from Thorlabs) was used. The light source was the same in all the measurements. The intensity of the reflected light when no curvature ($C = 0$ m$^{-1}$) was applied to the device was considered as $P$ and the changes caused by curvature as $\Delta P$. At $C = 0$ m$^{-1}$, the wavelength position of the peak reflection was considered to be $\lambda_m$ and $I_R = 1$.

Figures 6(a) and 6(b) show the spectra observed when the curvature at two perpendicular directions was applied to the device described in the above paragraph. Figures 6(c) and 6(d) show the averaged curvature sensitivities that were measured in seven different orientations of the MCF. The core orientations with respect to curvature are illustrated in Figs. 4 and 5. Note that when the wavelength shift is larger, the changes in intensity are minimal and vice versa. The different values of sensitivities at different orientations of the MCF cores with respect to curvature were expected due to the asymmetry of the device.

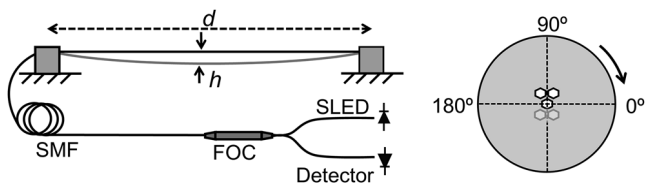

**FIG. 5.** Schematic diagram of the measuring setup and the sensor interrogation; $h$ is the deflection of the device and $d$ is the distance between the two supports. FOC is fiber optic coupler or circulator, SMF is single mode fiber, and SLED is superluminescent light emitting diode. The MCF core orientation with respect to the applied curvature is indicated.

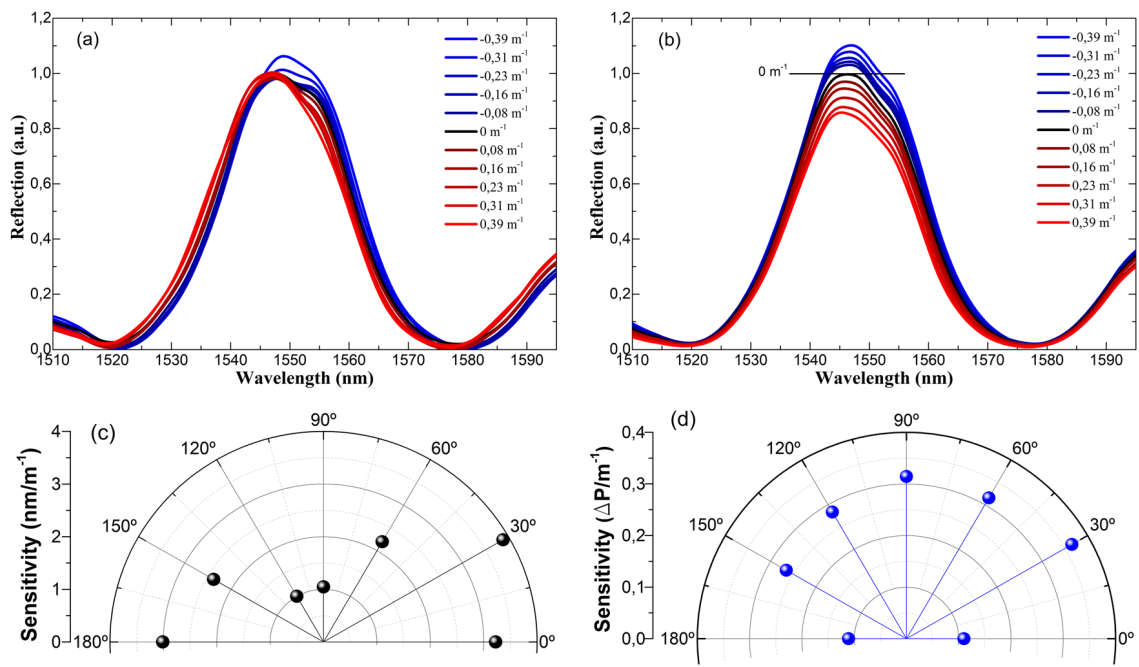

**FIG. 6.** [(a) and (b)] Reflection spectra at different curvatures observed when the position of the MCF was at 0° and 90°, respectively, according to Figs. 4(c) and 4(d). [(c) and (d)] Average curvature sensitivity measured by monitoring wavelength shift or intensity changes. In all cases, the MCF device had $L_1$ = 17.40 mm and $L_2$ = 18.20 mm.

The discrepancy between simulations and experimental results with regard to shifts of the spectra may be due to the strain induced to the device and curvature of the SMF–MCF junctions, as these are inevitable in an experiment. In addition, during the measurements, the two segments of MCF may not experience exactly the same curvature. In the simulations, however, the two stubs of MCF were supposed to be exclusively subjected to the same curvature. Nonetheless, regardless of the orientation of the MCFs with respect to curvature, the wavelength position and height of the reflection peak (intensity) can be simultaneously tracked. Hence, it is possible to know the direction and amplitude of the curvature applied to the device.

The drastic changes in the reflection spectrum of the SMF–MCF1–MCF2–SMF structure when it is subjected to curvature can be explained with Eq. (2) and with the simulations shown in Fig. 4. Note that the structure is composed of two MCF segments that are highly sensitive to bending. Moreover, the reflection spectrum results from the multiplication of two spectra that move in opposite directions. This causes the height of the resulting reflection peak to increase or decrease. Consequently, the total intensity detected by using the photodetector increases or decreases depending on the direction of curvature.

In real-world applications, fiber optic curvature sensors are attached or integrated to structures or devices. Thus, to investigate the performance of our curvature sensor in more detail, the sample described in Fig. 3 was glued on a thin rectangular plastic beam that was secured with two supports separated by a fixed distance. The orientation of the cores of the segments of MCF with respect to the plastic beam was approximately as that shown in Fig. 5. This means that a segment of MCF had two cores up and the other two cores down with respect to the direction of the curvature. Again, a translation stage with micrometer resolution was used to bend the beam upward (convex curvature) and downward (concave curvature) in a controlled manner. Other curvature orientations were not possible due to the geometry of the beam. The stage was located in the middle point of the distance between the two supports. The MCF1–MCF2 junction of the structure was located in the same position than translation stage.

Figure 7 summarizes the behavior of our sensor when it was subjected to concave and convex curvatures. Note that the shift of the spectrum is to longer wavelengths in the former case and to shorter wavelengths in the latter case. The figure also shows the calibration curve for concave and convex curvatures. It can be noted that the response of our device in both cases is linear. From the calibration curve, the curvature sensitivities were calculated to be 791 pm/m⁻¹ for concave curvature and 950 pm/m⁻¹ for convex curvature. The discrepancy in the values of sensitivities of our device can be attributed to imperfections of the same, for example, the MCFs may not be exactly 180° with respect to each other. Strain applied to the MCFs and curvature of the SMF–MCF segments may also induce shifts to the reflection spectra.

In Fig. 8, we show the observed changes in $\Delta P/P$ for different values of concave and convex curvatures. It can be noted that when the device was subjected to concave curvature, the value of $\Delta P/P$ decreased, and it increased when the curvature on it was convex. Note also that the value of $\Delta P/P$ reached the baseline ($\Delta P/P = 0$) when the curvature was removed from the sensor. The calibration curves for concave and convex curvatures are also shown in Fig. 8. The sensitivities for concave and convex curvature were found to be almost identical, 4.66 dB/m$^{-1}$, which is slightly higher than those of the intensity-modulated curvature sensors reported in Refs. 26, 37, and 38.

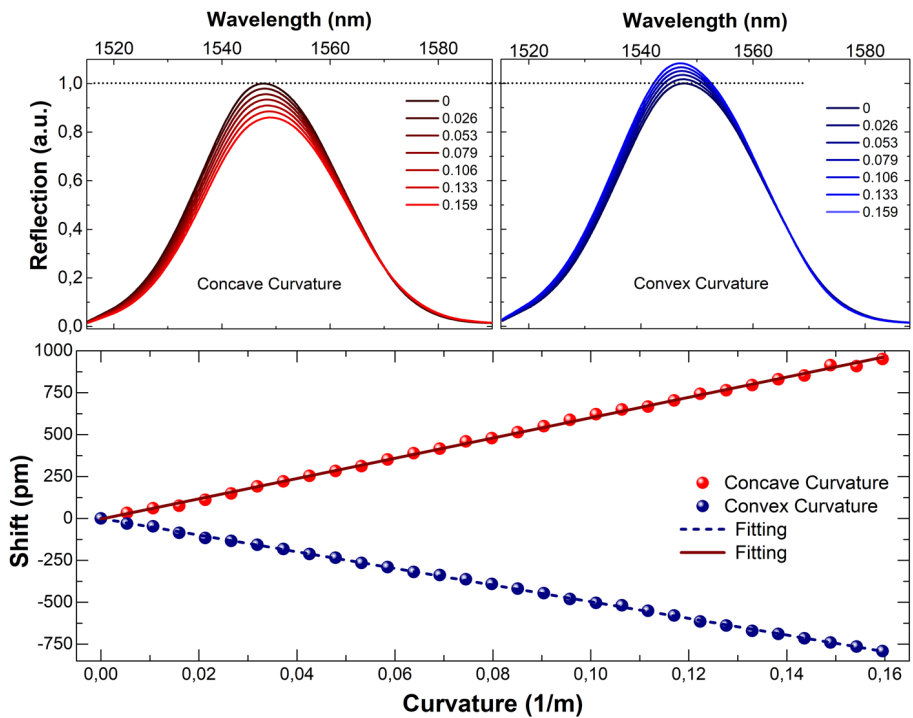

**FIG. 7**. Top: spectra observed when the beam shown in Fig. 5 was curved downward (left) and upward (right). The values of curvature (in m$^{-1}$) are indicated in the graphs. Bottom: calibration curves for concave and convex curvatures.

The results shown in Fig. 8 suggest that with our device and an inexpensive intensity-based interrogation system, it is possible to distinguish concave and convex curvatures as well as the amplitude of the applied curvature. If maximum sensitivity is needed in a particular curvature direction, the cores of the MCF can be oriented properly. We believe that these features cannot be achieved with other fiber optic curvature sensors reported so far in the literature.

## CONCLUSIONS

In this work, we have reported on a simple MCF curvature sensor that comprises two short segments of strongly coupled MCF fusion spliced and rotated with respect to each other. The fabrication of the device only involves cleaving and fusion splicing; such processes are well established in the fiber optics industry. The sensor can be interrogated with a low power SLED and a miniature spectrometer or a simple photodetector. It was found that the sensor behaves as two SMF–MCF–SMF structures in series and the reflection spectrum exhibited a single, narrow peak whose height and position in wavelength can be simultaneously determined with high accuracy.

The proposed device was assessed as a curvature sensor. It was found that for this application, it is able to provide the amplitude and the direction of curvature no matter how the cores of the MCF are oriented with respect to the direction of curvature. Moreover, our sensor can be interrogated in two different manners. When the sensor was subjected to concave curvature, the reflection spectrum shifted to red and the intensity decreased. However, when convex curvature was applied to the device, the shift was to blue and the intensity decreased.

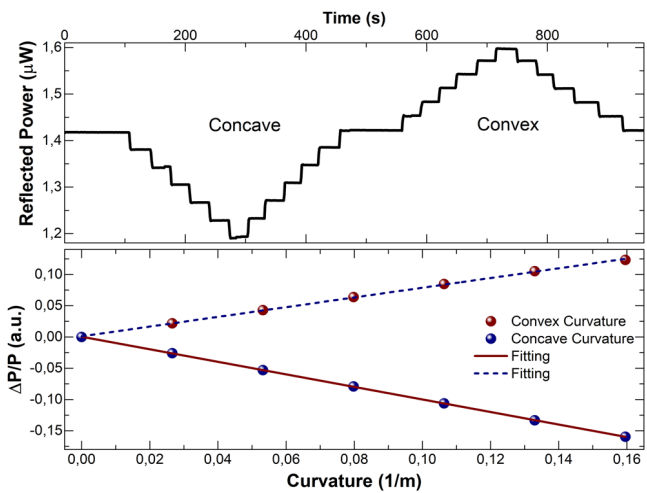

**FIG. 8**. Top: relative power changes as a function of time when the beam, hence the MCF segments, was bent downward (concave curvature) and upward (convex curvature). The step in each case is 0.0266 m$^{-1}$. Bottom: calibration curve.

The curvature sensitivity of the sensor reported here was found to be 4.66 dB/m$^{-1}$ when intensity changes were correlated with curvature. Such sensitivity can be sufficient in several applications.

We believe that the composed MCF structure reported here can be used for different sensing applications. Vibrations, for example, can be translated to periodic concave and convex curvatures on the device and hence to periodic intensity changes. It also seems possible to sense pressure or lateral force as they can induce curvature to the MCF segments. Therefore, cost effective, highly sensitive force, pressure, or vibration (accelerometers) sensors can be devised with the platform proposed here.

## ACKNOWLEDGMENTS

The authors acknowledge the financial support of the Spanish MINECO under Project Nos. PGC2018-101997-B-I00 and RTI2018-094669-B-C31 of the Eusko Jaurlaritza (Basque Government) under Project Nos. IT933-16 and ELKARTEK.

## DATA AVAILABILITY

The data that support the findings of this study are available from the corresponding author upon reasonable request.

## REFERENCES

[1] K. Saitoh and S. Matsuo, J. Lightwave Technol. **34**, 55–66 (2016).

[2] E. M. Dianov, S. L. Semjonov, and I. A. Bufetov, Quant. Electron. **46**, 1 (2016).

[3] C. Xia, M. A. Eftekhar, R. A. Correa, J. E. Antonio-Lopez, A. Schülzgen, D. Christodoulides, and G. Li, IEEE J. Sel. Top. Quantum Electron. **22**, 196–207 (2015).

[4] S. Sivankutty, V. Tsvirkun, G. Bouwmans, D. Kogan, D. Oron, E. R. Andresen, and H. Rigneault, Opt. Lett. **41**, 3531–3534 (2016).

[5] R. G. Duncan, M. E. Froggatt, S. T. Kreger, R. J. Seeley, D. K. Gifford, A. K. Sang, and M. S. Wolfe, Proc. SPIE **6530**, 65301S (2007).

[6] J. P. Moore and M. D. Rogge, Opt. Express **20**, 2967–2973 (2012).

[7] J. Villatoro, E. Antonio-Lopez, A. Schülzgen, and R. Amezcua-Correa, Opt. Lett. **42**, 2022–2025 (2017).

[8] J. Villatoro, A. Van Newkirk, E. Antonio-Lopez, J. Zubia, A. Schülzgen, and R. Amezcua-Correa, Opt. Lett. **41**, 832–835 (2016).

[9] D. Monzon-Hernandez, A. Martinez-Rios, I. Torres-Gomez, and G. Salceda-Delgado, Opt. Lett. **36**, 4380–4382 (2011).

[10] Q. Wang and Y. Liu, Measurement **130**, 161–176 (2018).

[11] J. A. Martin-Vela, J. M. Sierra-Hernandez, A. Martinez-Rios, J. M. Estudillo-Ayala, E. Gallegos-Arellano, D. Toral-Acosta, T. E. Porraz-Culebro, and D. Jauregui-Vazquez, IEEE Photonics Technol. Lett. **31**, 1265–1268 (2019).

[12] Y.-P. Wang and Y.-J. Rao, IEEE Sens. J. **5**, 839–843 (2005).

[13] D. Z. Stupar, J. S. Bajic, L. M. Manojlovic, M. P. Slankamenac, A. V. Joza, and M. B. Zivanov, IEEE Sens. J. **12**, 3424–3431 (2012).

[14] M. Jang, J. S. Kim, S. H. Um, S. Yang, and J. Kim, Opt. Express **27**, 2074–2084 (2019).

[15] L. Yuan, J. Yang, Z. Liu, and J. Sun, Opt. Lett. **31**, 2692–2694 (2006).

[16] H. Qu, G. F. Yan, and M. Skorobogatiy, Opt. Lett. **39**, 4835–4838 (2014).

[17] C. Li, T. Ning, C. Zhang, J. Li, C. Zhang, X. Wen, H. Lin, and L. Pei, Sens. Actuators, A **248**, 148–154 (2016).

[18] S. Zhang, A. Zhou, H. Guo, Y. Zhao, and L. Yuan, OSA Continuum **2**, 1953–1963 (2019).

[19] W. Chen, Z. Chen, Y. Qiu, L. Kong, H. Lin, C. Jia, H. Chen, and H. Li, Appl. Opt. **58**, 8776–8784 (2019).

[20] J. R. Guzman-Sepulveda and D. A. May-Arrioja, Opt. Express **21**, 11853–11861 (2013).

[21] G. Salceda-Delgado, A. Van Newkirk, J. E. Antonio-Lopez, A. Martinez-Rios, A. Schülzgen, and R. Amezcua Correa, Opt. Lett. **40**, 1468–1471 (2015).

[22] A. V. Newkirk, J. E. Antonio-Lopez, A. Velazquez-Benitez, J. Albert, R. Amezcua-Correa, and A. Schülzgen, Opt. Lett. **40**, 5188–5191 (2015).

[23] G. M. H. Flockhart, W. N. MacPherson, J. S. Barton, J. D. C. Jones, L. Zhang, and I. Bennion, Opt. Lett. **28**, 387–389 (2003).

[24] D. Barrera, I. Gasulla, and S. Sales, J. Lightwave Technol. **33**, 2445–2450 (2014).

[25] D. Zheng, J. Madrigal, H. Chen, D. Barrera, and S. Sales, Opt. Lett. **42**, 3710–3713 (2017).

[26] P. Saffari, T. Allsop, A. Adebayo, D. Webb, R. Haynes, and M. M. Roth, Opt. Lett. **39**, 3508–3511 (2014).

[27] S. Wang, W. Zhang, L. Chen, Y. Zhang, P. Geng, Y. Zhang, T. Yan, L. Yu, W. Hu, and Y. Li, Opt. Lett. **42**, 4938–4941 (2017).

[28] D. Barrera, J. Madrigal, and S. Sales, J. Lightwave Technol. **36**, 1063–1068 (2018).

[29] A. W. Snyder, J. Opt. Soc. Am. **62**, 1267–1277 (1972).

[30] N. Kishi and E. Yamashita, IEEE Trans. Microwave Theory Tech. **36**, 1861–1868 (1988).

[31] J. Hudgings, L. Molter, and M. Dutta, IEEE J. Quantum Electron. **36**, 1438–1444 (2000).

[32] A. Perez-Leija, J. Hernandez-Herrejon, H. Moya-Cessa, A. Szameit, and D. N. Christodoulides, Phys. Rev. A **87**, 013842 (2013).

[33] R. P. Murphy, S. W. James, and R. P. Tatam, J. Lightwave Technol. **25**, 825–829 (2007).

[34] A. Varguez-Flores, G. Beltran-Perez, S. Munoz-Aguirre, and J. Castillo-Mixcoatl, J. Lightwave Technol. **27**, 5365–5369 (2009).

[35] D. Barrera, J. Villatoro, V. P. Finazzi, G. A. Cárdenas-Sevilla, V. P. Minkovich, S. Sales, and V. Pruneri, J. Lightwave Technol. **28**, 3542–3547 (2010).

[36] H. Liao, P. Lu, X. Fu, X. Jiang, W. Ni, D. Liu, and J. Zhang, Opt. Express **25**, 26898–26909 (2017).

[37] Y. Fu, H. Di, and R. Liu, Opt. Laser Technol. **42**, 594–599 (2010).

[38] J. Shi, F. Yang, D. Yan, D. Xu, C. Guo, H. Bai, W. Xu, Y. Wu, J. Bai, S. Zhang, T. Liu, and J. Yao, Opt. Express **27**, 23585–23592 (2019).